\documentclass{emulateapj}
\input{epsf.sty}

\newcommand{\kms}{km\thinspace s$^{-1}$}     % km s-1
\newcommand{\arcs}{\ifmmode {'' }\else $'' $\fi}  % arc se
\newcommand{\arcm}{\ifmmode {' }\else $' $\fi}    % arc min
\newcommand{\mstar}{\ifmmode {M_{HI_\ast}}\else $M_{HI_\ast}$\fi}
\newcommand{\msolar}{M$_\odot$}
\newcommand{\gapprox}{\ifmmode \buildrel > \over {_\sim} \else $\buildrel >\over {_\sim}$\fi}
\newcommand{\lapprox}{\ifmmode \buildrel < \over {_\sim} \else $\buildrel <\over {_\sim}$\fi}
\newcommand{\lj}{L$_J$ }
\newcommand{\mhi}{M$_{HI}$ }

\begin{document}

\shortauthors{Rosenberg, Schneider, \& Posson-Brown}
\shorttitle{Gas and Stars in an HI-Selected Sample}

\title{Gas and Stars in an HI-Selected Galaxy Sample}

\author{Jessica L. Rosenberg\altaffilmark{1}, \altaffilmark{2}}
\affil{Center for Astrophysics \& Space Astronomy, Department of Astrophysical
and Planetary Sciences, University of Colorado, Boulder, CO 80309}
\altaffiltext{1}{National Science Foundation Astronomy and Astrophysics Postdoctoral
Fellow}
\author{Stephen E. Schneider} 
\affil{Department of Astronomy, University of Massachusetts, Amherst, MA 01003}
\author{Jennifer Posson-Brown\altaffilmark{2}}
\affil{Smith College, Northampton, MA 01063}
\altaffiltext{2}{Present address: Center for Astrophysics, Cambridge, MA 02138}

\begin{abstract}

We present the results of a J-band study of the HI-selected Arecibo Dual-Beam
Survey and Arecibo Slice Survey galaxy samples using the 2 Micron All-Sky Survey 
data. We find that these galaxies span a wide range of stellar and gas
properties. However, despite the diversity within the samples, we find a very tight
correlation between luminosity and size in the J-band, similar to that we previously
found (Rosenberg \& Schneider 2003) between the HI mass and size. We also find
that the correlation between the baryonic mass and the J-band diameter is even
tighter than between the baryonic mass and the rotational velocity.

\end{abstract}


\section{Introduction}
\label{sec:2mass_intro}

In recent years several large ``blind" 21 cm surveys for galaxies have been
conducted providing a measurement of the gas in galaxies in the local universe 
(Meyer et al. 2004; Rosenberg \& Schneider 2000;
Zwaan et al. 1997, Spitzak \& Schneider 1998). These surveys have revealed
many galaxies with unusual characteristics: some without a definite optical
counterpart and others with very high HI to stellar mass ratios, for example. These
galaxy samples give us an opportunity to explore how the stellar 
properties of gas-rich galaxies differ from those of optically luminous
systems. 

Our understanding of the relationship between the global properties of gas and 
stars in galaxies has mostly
been driven by studies of optically-selected, high surface brightness 
galaxies (e.g. Scodeggio \& Gavazzi 1993; Huchtmeier \& Richter 1985; Fisher 
\& Tully 1981). Various efforts have been made to extend these studies to lower 
surface brightness galaxies, (e.g., Galaz et al. 2002, McGaugh et al. 2000, 
O'Neil et al. 2000, Sprayberry et al. 1995) revealing a great diversity of properties 
outside of the traditional ``norms"  defined by the high surface brightness samples, 
but such surveys remain tied to the requirement that
the galaxies have formed stars in sufficient numbers and surface densities to be
detected optically. Extragalactic HI surveys provide one of the
few ways to probe the galaxy population independent of their
luminosity and surface brightness. HI galaxy selection is also complementary to
optical galaxy selection with respect to star formation history since the conversion 
of gas to stars renders a galaxy more visible optically, but less visible in an HI 
survey.

The blind HI surveys conducted with the Arecibo radio telescope remain among the 
deepest to
date sampling galaxies with HI fluxes almost an order of magnitude smaller than
the very large, but shallow HI Parkes All-Sky Survey (HIPASS, Meyer et al. 2004) 
and probing
the lowest mass galaxies over a wider range of environments. 
The stellar properties of the Arecibo samples of Zwaan et al (1997) and Spitzak \& 
Schneider 
(1998) have been studied at optical wavelengths. These studies have provided a
picture of the gas-rich galaxies in the universe as containing a sub-sample of
low-luminosity and low-surface-brightness galaxies (Spitzak \& Schneider 1998) in
higher proportion than found in optical surveys.

Infrared observations more nearly reflect the total stellar mass of galaxies, since
they are less sensitive to the star formation rate and history and are less affected 
by dust than optical observations. In 
this paper we use the data from the 2 Micron All-Sky Survey (2MASS) to study the 
infrared properties of galaxies from the Arecibo Dual-Beam Survey (Rosenberg \& 
Schneider 2000; RS) and from the Slice Survey (Spitzak \& Schneider 1998; SS). 
Both of these surveys are ``blind" HI surveys that probed the gas-rich galaxy 
populations in the local universe. The 2MASS observations of this galaxy sample 
are useful for examining the relationship between gas and stars in gas-rich 
galaxies in the local universe. While 2MASS provides information about the stellar 
mass in these galaxies, it suffers from a lack of
surface brightness sensitivity which limits the galaxy detection rate. Nevertheless, 
over 85\% of the galaxies in each sample were detected.

We discuss the HI and 2MASS data used in this study in \S 2 along with details
about how some of the galaxies were measured from the 2MASS images. In \S 3 we
discuss the relationship between the gaseous and stellar properties of the
galaxies and discuss the surface densities of gas and stars in \S 4. In \S 5 we
discuss the 2MASS images of the galaxies and in \S 6 we summarize our results.

\section{DATA}
\label{sec:2mass_samples}

\subsection{Sample Selection - The Arecibo Dual-Beam and Slice Surveys}

We examine the stellar properties of two HI-selected galaxy 
samples. Both the RS and SS surveys are ``blind" 21 cm surveys carried out with 
the Arecibo 305 m telescope prior to the Gregorian upgrade. We have used these
surveys to identify galaxies purely based on their gas content out to 7977 \kms\ 
and 8340 \kms\ respectively. The RS survey covered $\sim$ 430 deg$^2$ in the 
main beam and
detected 265 galaxies while the SS survey covered $\sim$ 55 deg$^{2}$ and
detected 75 galaxies. The selection functions and HI mass functions have 
been studied in detail and are presented in Rosenberg \& Schneider (2002) 
and Schneider, Spitzak, \& Rosenberg (1999) for the RS and SS surveys 
respectively. 

The details of the calculations of line width, velocity,
distance, and HI mass for the entire sample are presented in RS. The SS survey 
additionally includes broadband B, R and I optical data. There are three galaxies 
from the RS survey for which the RS line widths
differed from the literature values because one horn of the velocity profile was
missed in the survey measurement. In these cases (RS 17, RS 109, and RS 189) we use 
the values from the Huchtmeier \& Richter catalog (1985) when using line widths to
calculate the dynamical mass (see \S3). 

\subsection{The 2-Micron All Sky Survey Data}

The 2 Micron All-Sky Survey (2MASS) provides simultaneous J, H, 
and K$_s$-band observations of the entire sky; the 2MASS project processed the
data to generate a point source catalog and an extended source catalog. We use 
the extended source catalog data for the RS and SS galaxies in the following
analyses and also supplement this with our own analysis of the images for a number of
galaxies that were undetected by the standard processing procedures (discussed
in \S 2.2.1). A full 
description of the galaxy detection algorithm and the resulting
catalogs are available in the Explanatory Supplement to the 2MASS All Sky Data 
Release\footnote{http://www.ipac.caltech.edu/2mass/releases/allsky/doc/explsup.html
by Cutri, R.M., Skrutskie, M.F., Van Dyk, S., Beichman, C.A., Carpenter, J.M., 
Chester, T., Cambresy, L., Evans, T., Fowler, J., Gizis, J., Howard, E., Huchra, 
J., Jarrett, T., Kopan, E.L., Kirkpatrick, J.D., Light, R.M, Marsh, K.A., 
McCallon, H., Schneider, S., Stiening, R., Sykes, M., Weinberg, M., Wheaton, W.A., 
Wheelock, S., Zacharias, N.} (Cutri et al.). 

The 2MASS extended source catalog has undergone several iterations and we use
results from both the Version 2 and Version 3 catalogs in these analyses. We use 
Version 2 in addition to Version 3 because this earlier version of the software 
used previous galaxy 
catalog positions to seed the search algorithms (in addition to doing 
an independent automated search) and therefore measured a number of fainter 
sources that Version 3 did not detect, since it used only the automated
algorithms. In addition, there are some large galaxies for which we report the
Version 2 values rather than the Version 3 values because some of the
information if missing the the version 3 catalog.

The J-band observations are the most sensitive of the three 2MASS bands and our 
HI-selected sources tend to be blue on average, so we use these data in our 
analyses.
The catalog data were used for 45 (4 from Version 2, 41 from Version 3) of the 75 
SS galaxies and 180 (50 from Version 2, 130 from Version 3) of the 265 RS 
galaxies. Some of the galaxies that were not in the extended source catalogs
can be identified on the full resolution 2MASS images; in these cases we measured 
the galaxy from the image. Details of our image analysis are described below. 
There were an additional 19 SS and 47 RS galaxies that we measured from the 
images. 
For 11 of the SS galaxies and 38 of the RS galaxies there was no cataloged
detection and we were not able to measure the galaxy from the images. For
most of the galaxies that were not measured, the source was just too faint to be 
detected in the short 2MASS exposures, but some exceptions are described in the
specific galaxy notes below. 

The near infrared measurements for the galaxies in both the RS and SS surveys
are given in Tables 1 and 2 in the Appendix. The tables also list the source for
the measurements, i.e., whether the data come from Version 2 or 3 of the 2MASS
catalog, were measured from the image, or if the source was not detected. 

\subsubsection{2MASS Measured Data}

To measure the galaxies on the 2MASS images we have used the ELLIPSE package in 
IRAF. The brighter galaxies were interactively fit with ellipses such that the
ellipse parameters were allowed to change in the highest surface brightness
regions but when the fits became uncertain in the outer regions, the ellipse 
parameters were held fixed while additional steps in radius were taken. For the 
lowest surface brightness galaxies, even the central regions had uncertain 
parameter fits, so only fixed circular apertures were used for the photometry. 

ELLIPSE was run at least 
twice for each galaxy so that the outer ``background" ellipses could be used for 
background subtraction. Then the fitting was rerun on the background-subtracted 
image.  The background subtraction was repeated if an adequately flat background 
had not been obtained on the first pass. In all cases, the point sources in the 
image were masked prior to the ellipse fitting.

\subsubsection{Notes About Selected Galaxies}

We provide information about the near infrared measurements for a few 
galaxies that warrant a comment. The numbers refer to the entry number in the RS 
survey catalog (see Table 1 for the ADBS names).

\begin{itemize}
\item{{\bf RS 50}: The HI detection of this system includes an interacting pair 
of galaxies. Since we do not separate the pair in the HI observation we do not 
try to correlate the HI with a near infrared measurement of one galaxy or the 
other.}
\item{{\bf RS 73}: We do not report an infrared measurement for this galaxy 
because it is too contaminated by a bright star.}
\item{{\bf RS 79}: We use the Version 2 measurement of the brighter galaxy in the 
interacting pair. Both galaxies are probably contributing to the HI, but 
including the fainter of the two galaxies would change the J-band magnitude by 
less than 0.3 mag.}
\item{{\bf RS 184}: This galaxy is in the Version 3 extended source catalog.
However, we use measurements from the image because the center of this very low 
surface brightness galaxy was missed in favor of a field star by the galaxy 
detection algorithm.}

\end{itemize}

\subsection{Using the 2MASS data as a Measure of Stellar Mass}

As a basis for understanding the properties of these HI-selected galaxies we
want to compare the gas mass and the stellar mass in these systems. The 2MASS
data are particularly good for measuring the stellar mass as they are more
sensitive to low mass stars and less sensitive to dust and star formation than
optical observations. However, there is substantial scatter in the relationship 
between luminosity and stellar mass even in the infrared. The scatter can be
decreased by using information about the galaxy's color (Bell \& de Jong 2001).

Based on stellar population synthesis models, a variety of estimates have been 
made for various stellar initial mass functions and star-formation histories. 
Bell \& de Jong (2001) present a range of models that yield realistic colors and 
surface brightnesses for present-day galaxies. Bell (private communication) has 
provided us with J-band values for these same models and finds a tight 
correlation between (B-R) color and the ratio of stellar mass to J-band 
luminosity:

\begin{equation}
log(M_{star}/L_J) = 0.552 (B-R) - 0.724
\end{equation}

The value of $M_{star}/L_J$ varies from $\sim0.4$ for blue 
galaxies [$(B-R)\sim0.6$] to $\sim1.2$ for red galaxies [$(B-R)\sim1.6$]. 
Including differences between models and measurement uncertainties, there is
thus a range of almost a factor of 4 in the relationship between 
stellar mass and J-band luminosity for the normal range of galaxy colors.
When the color is known, the total range reduces to about a factor of 1.5.
These results appear to be consistent with alternative models carried 
out for I and H bands (cf. McGaugh \& de Blok 1997;  McGaugh et al. 2000).

\begin{figure}[ht]
\plotone{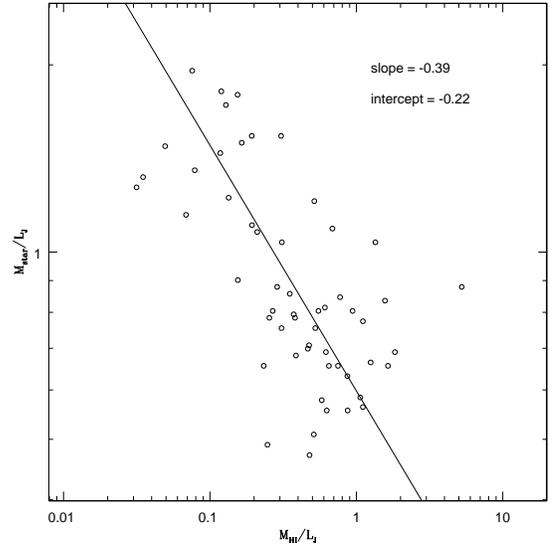}
\caption{The relationship between the mass-to-light ratio calculated using
Equation 1 and the M$_{HI}$/L$_J$ for the SS data. The fit to these data are
used to improve the estimate of M$_{star}$/L$_J$ for the data for which B-R
measurements are not available.}
\label{fig:MLcomp2}
\end{figure}

The SS galaxies were previously observed in broadband colors, which allows us to 
estimate their mass-to-light ratios more precisely. Their average (B-R) 
color was 1.0, so their average M$_{star}$/L$_J$ value is 0.7. Figure
\ref{fig:MLcomp2} shows the relationship between M$_{star}$/L$_J$ for the SS
galaxies calculated using Equation 1 and their ratio of HI mass to J-band luminosity,
M$_{HI}$/L$_J$. The total neutral hydrogen mass was calculated from the published
data using the well-known relationship M$_{HI}=2.36\times10^5 D^2 I_{HI}$, where $D$
is the distance in Mpc and $I_{HI}$ is the integrated flux in Jy km s$^{-1}$. 
The standard deviation around the line is 0.15 while the
standard deviation around the average value of $log(M_{star}/L_J$) = -0.15 is 0.19. 
The correlation is not extremely tight, but it allows us to improve our estimate 
of the mass-to-light-ratios for the galaxies
that do not have broadband colors: the RS galaxies and the faintest SS
galaxies.

In addition to the intrinsic and measurement uncertainties in determining 
M$_{star}$/L$_J$, an additional source of
scatter is introduced in the determination of stellar mass because the 2MASS 
data are not very deep. Uncertainties in the J-band flux are caused by the difficulty 
in measuring the isophotal sizes near the limit of the survey surface brightness
sensitivity, particularly for the lowest surface brightness galaxies. We compare the 
isophotal measurements of galaxy magnitudes at 21 mag arcsec$^{-2}$ to the
measurements in the largest aperture before the noise takes over (i.e., before the 
values start oscillating between fainter and brighter values due to noise). For 
the 2MASS cataloged values, we use the brightest aperture value reported. To be 
consistent with the catalog measurements, we restrict our own measured apertures 
to those used in the catalog (ie., 5, 7, 10, 15, 20, 25, 30, 40, 50, 60, or 70
arcsec). However, 19 of the RS galaxies are significantly larger than 70$\arcsec$ so
the largest aperture is not a good measurement of its size of flux. For these
galaxies, we use the extrapolated J-band magnitude so that the size and flux are not
severely underestimated. Table 1 indicates for which galaxies the extrapolated
values are used. There are no SS galaxies that are significantly larger than
the 70$\arcsec$ aperture.

\begin{figure}[ht]
\plotone{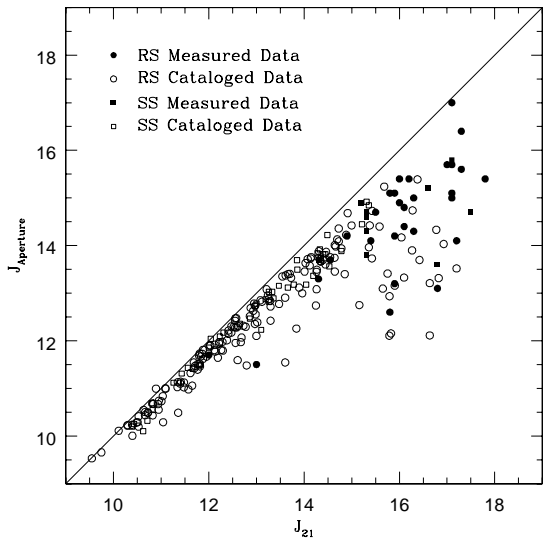}
\caption{The relationship between aperture and 21 mag arcsec$^{-2}$ isophotal
magnitudes from the RS and the SS samples. The open symbols indicate magnitudes 
that were obtained from the 2MASS extended source catalog, Version 2 or 3. The 
closed symbols indicate magnitudes that were measured from the images.}
\label{fig:magcomp}
\end{figure}

Figure \ref{fig:magcomp} shows that the magnitudes for most of the faint
galaxies are severely underestimated if the isophotal value is used. 
For all of the J-band luminosities in this paper we use the aperture
measurement values. Additionally, we adopt a value of \msolar(J) = 3.73 (Johnson 1966; 
Allen 1973) for our conversion to solar luminosities. 

\section{Stars and Gas in an HI-Selected Sample}

By using the RS and SS data, we have selected a wide range of galaxy types with
the only criteria being that they contain neutral hydrogen. These systems span 
the range from
dwarfs and irregulars to early-type spirals and interacting systems. Appendix A
shows the 2MASS near infrared images for these galaxies; their morphology is
discussed in greater detail in \S 5.

\begin{figure}[ht]
\plotone{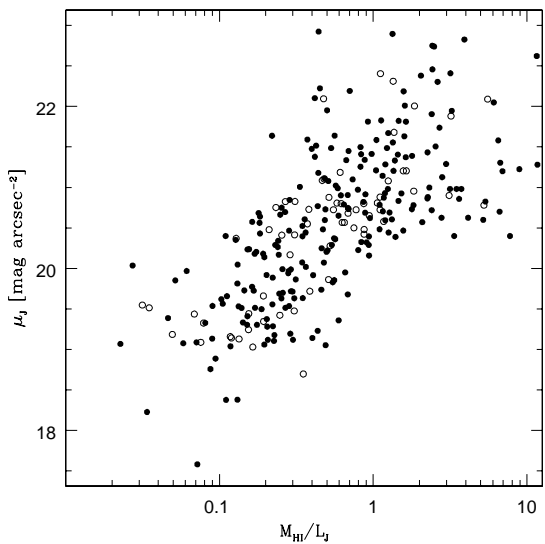}
\caption{The relationship between the average J-band surface brightness and the
ratio of HI mass to J-band luminosity for the RS galaxies (filled circles) and
the SS galaxies (open circles).}
\label{fig:sbcomp}
\end{figure}

Figure \ref{fig:sbcomp} shows the relationship between the J-band surface
brightness and M$_{HI}$/L$_J$ for the RS galaxies (filled circles) and for the
SS galaxies (open circles). The J-band surface brightness is determined within
the aperture radius (or extrapolated radius for the largest galaxies) determined 
as discussed in the previous section. The surface brightness that we derive is, 
roughly, a surface brightness at a slightly fainter isophote than J=21 mag 
arcsec$^{-2}$. This figure illustrates that, in general, the gas-rich galaxies
(galaxies with high values of M$_{HI}$/L$_J$) are the lower surface brightness
ones indicating that HI-selection is a good way to find low surface brightness
galaxies. Despite not having a very precise measure of surface brightness, the
correlation with M$_{HI}$/L$_J$ is pretty good.

The typical (B-J) color for a spiral galaxy is between 2 and 3. Using (B-J) = 2 
as a conservative value, the average surface brightnesses of these HI selected
galaxies range between $\Sigma_J$ = 18 to 25 mag arcsec$^{-2}$. There are a variety of
different definitions of what constitutes a low surface brightness galaxy. One 
definition refers only to the central surface brightness of the disk component 
after a bulge-disk decomposition has been carried out. By this definition, many
galaxies with high surface brightness bulges are categorized as LSB, because the
disk component is faint. Another common definition uses the mean blue surface 
brightness within the $\mu_{B_0} = $ 25.0 mag arcsec$^{-2}$ isophote, giving the
label of LSB to galaxies with inclination-corrected mean surface brightnesses dimmer 
than $\langle\mu_{B_0}\rangle > $25.0 mag arcsec$^{-2}$.

Since we cannot perform bulge-disk decompositions with the present data, our mean
J-band surface brightnesses within the $\mu_{J} = $ 21.0 mag arcsec$^{-2}$ are most
readily comparable to the latter definition, although for typical (B-J) colors we
measure the mean within a brighter isophote. It appears, though, that analogous
to the B-band definition of LSB, we might call galaxies dimmer than 
$\langle\mu_{J}\rangle > $21.0 mag arcsec$^{-2}$ infrared
LSBs. This cutoff for LSBs also marks the approximate point at which the total
stellar luminosity becomes smaller than the HI mass (in solar units).

The total baryonic mass was estimated from the HI and J-band emission using:
\begin{equation}
M_{bar} = M_{star} + 1.4\times M_{HI} .
\end{equation}
where M$_{star}$ is the total stellar mass, derived from the J-band luminosity
mass-to-light ratio discussed in the previous section, and the total gas mass is
estimated from M$_{HI}$ multiplied by 1.4 to account for helium and metals.
Obviously, including an estimate of the molecular and ionized gas would increase 
this further. The stellar and gas masses both have significant uncertainties 
associated with them (e.g., due to the metallicity of the stars and the amount 
of molecular gas present), but the effect of internal extinction on the estimate 
is small compared to its effect at optical wavelengths since the extinction at J is only
about 20\% of that at B. Furthermore,
the additional gas mass is likely to be largest for earlier type galaxies 
(Young \& Knezek 1991), which are dominated by their stellar content, and the 
uncertainties in the stellar mass are largest for the dim galaxies that are 
generally dominated by their neutral hydrogen mass, so the uncertainties should
have little impact on the total baryonic mass estimates.

\begin{figure}[ht]
\plotone{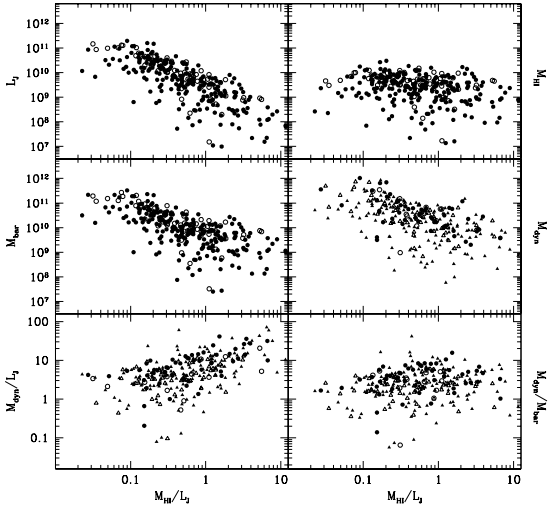}
\caption{The dependence of stellar luminosity, HI mass, total baryonic mass, and 
dynamical mass upon the ratio of gas to stars. The distance
independent quantities of the ratio of dynamical mass to J-band luminosity and
the ratio of dynamical mass to baryonic mass are also plotted as a function of
the gas to stars ratio. The filled circles are the RS galaxies, the open 
circles are the SS galaxies. The triangles plotted in panels containing
dynamical mass are the galaxies for which b/a$>$0.4 for which the
dynamical mass measurements may not be reliable.}
\label{fig:mlcomp}
\end{figure}

Figure \ref{fig:mlcomp} shows how the various measured and derived values of mass
and luminosity vary with M$_{HI}$/L$_{J}$. We find that the HI mass is essentially 
uncorrelated with the ratio for our HI-selected sample; consequently, the stellar 
luminosity declines roughly in inverse proportion to the M$_{HI}$/L$_{J}$ ratio. 
An indication of dynamical mass was calculated for our galaxies using their HI 
line widths (see RS and SS for a discussion of the measurements of line width) 
and optical dimensions to estimate the inclination and rotation speed, using the 
equation:

\begin{equation}
M_{dyn} = v_{rot}^2 r_{opt}/G .
\end{equation}

$v_{rot}$ is the inclination corrected line width ($v$/sin[i]). We limited the 
inclination correction to inclinations greater than $i>66^\circ$ (b/a $\le$ 0.4). 
This calculation of dynamical mass will not accurately estimate the masses of 
dwarf systems that are not primarily rotationally supported, and this 
mass estimate is directly dependent on the limits to which starlight is seen, 
but will, nevertheless give an indication of the total mass inside the faintest 
detected isophote. We have chosen to use the optical size in this calculation,
despite the fact that there is often significant HI mass outside of this radius,
because HI sizes are only available for a small subset of the RS data and are
not available for any of the SS data. The optical sizes of the RS galaxies are
derived from Palomar Sky Survey images and represent roughly the B-band 25 mag
arcsec$^{-2}$ isophotal size. In Figure \ref{fig:mlcomp} the galaxies 
with b/a$>$0.4 (more face-on systems for which the determination of dynamical 
mass is much less reliable) have been plotted as triangles.

Even though there are some clear trends in the properties of the galaxies in 
these surveys, Figure \ref{fig:mlcomp} also illustrates an important diversity. 
The width 
of the distribution of masses, luminosities and mass-to-light ratios at any 
given value of M$_{HI}$/L$_{J}$ is many times larger than might be caused by 
measurement uncertainties alone. The spread is probably even larger than these 
figures show since the lower-mass sources were only detectable to 
a limited distance within our search volume. What Figure \ref{fig:mlcomp} shows 
is that galaxies of any given HI mass span the entire range of gas-to-star 
ratios -- we see galaxies that are low in M$_{HI}$ because they
have turned most of their gas into stars and some that have low M$_{HI}$ 
because they are low mass systems with little of their mass in stars. 

We note the scarcity of galaxies with M$_{HI}$/L$_{J}$ \lapprox 0.03 in Figure
\ref{fig:mlcomp}. It is not clear whether this scarcity is a selection effect 
or a true deficit of galaxies. Presumably, 
some ellipticals and lenticulars would fall in this range, and some 
early-type systems {\it were} detected in these surveys. 
The total contribution of early type galaxies containing HI is uncertain. 
At such low M$_{HI}$/L$_{J}$ values, the HI selection criteria only permit us to 
detect high stellar luminosity sources, and even those only out to a relatively 
small distances. HI data for an unbiased optical sample is needed to determine 
their true statistical contribution.

\begin{figure}
\plotone{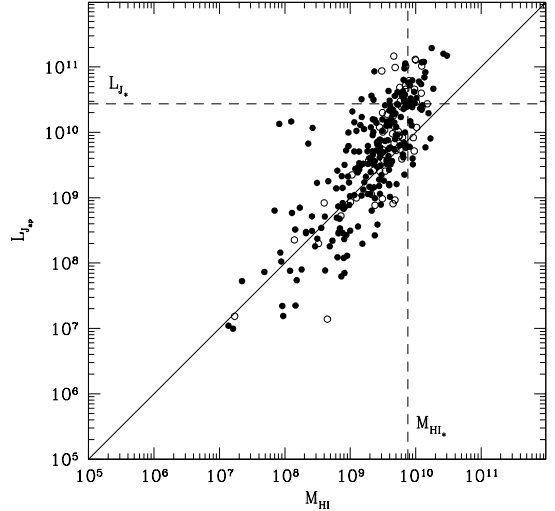}
\caption{The relationship between the J-band luminosity and HI mass for the RS
(filled circles) galaxies and the SS (open circles) galaxies. The dashed lines
indicate M$_{HI_{\ast}}$ (Rosenberg \& Schneider 2002) and L$_{J_{\ast}}$
(from the 2dF sample using 2MASS measurements; Cole et al. 2001). The solid 
line indicates a one-to-one relationship between the parameters.}
\label{fig:lcomp}
\end{figure}

The large range of galaxy properties and M$_{HI}$/L$_{J}$ values for the RS
and SS surveys indicates that there is not an easy conversion
between the HI mass function and the J-band luminosity function. Figure 
\ref{fig:lcomp} further demonstrates the problem with using one of these
parameters to predict the other. It is not clear that there is a linear
relationship between \lj\ and \mhi\ at all masses and luminosities and the 
spread around the correlation between the quantities is large.
As we noted in our earlier paper (Rosenberg \& Schneider 2002), at a given HI 
mass, there may be several orders of magnitude variation in the J-band 
luminosity, making HI mass a poor predictor of stellar content, and vice versa.

Both the baryonic and dynamical mass estimates are higher for sources with smaller
ratios of M$_{HI}$/L$_{J}$. 
As we show in Figure \ref{fig:mlcomp}, however, the dynamical mass-to-light 
ratio is higher on average for the gas-rich galaxies. This figure also shows
that the ratio of dynamical mass to baryonic mass is relatively flat as function
of M$_{HI}$/L$_{J}$ with a median value of 3.8 for galaxies with b/a $<$ 0.4.
The range in values that that we find for M$_{dyn}$/M$_{bar}$ is comparable to
the range for disk galaxies (Zavala et al. 2003).
Even excluding the more face-on systems for which the dynamical mass calculation
is problematic, there are 5 galaxies with dynamical masses that are smaller than
their baryonic mass estimate. For 2 of the RS cases, the measurement
of inclination is for the bright central region, but there is a more diffuse
light distribution that may indicate that these systems are more face on. For
the other 3 systems it is not clear why the dynamic mass is much
smaller than the baryonic mass except that they seem to have low rotation
velocities for such edge-on systems that might indicate that the line width has
not been properly measured (e.g., missing one horn of a double horned HI profile). 

\begin{figure}
\plotone{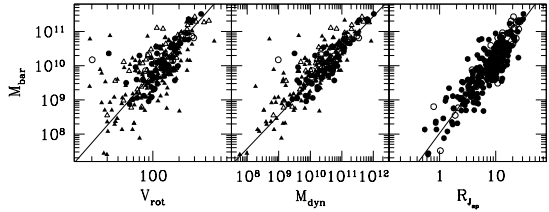}
\caption{The baryonic mass for the ADBS galaxies (filled circles) and the AS
galaxies (open circles) plotted against the dynamic mass, rotational velocity,
and J-band size. The triangles in the dynamic mass and rotational
velocity plots indicate the galaxies for which b/a$\ge$0.4 and are, therefore,
less reliable. The solid lines in the plots 
show the average of forward and backward least squares fits to the data
excluding the galaxies with b/a$\ge$0.4.}
\label{fig:mrcomp}
\end{figure}

Figure \ref{fig:mrcomp} shows the relationship between our estimate of the 
baryonic mass and a variety of other measurements that relate to the overall 
mass of a galaxy: the dynamical mass (as described in \S 3), the rotation speed, 
and the radius. Note that we have marked the face-on (b/a$\ge$0.4) galaxies with
triangles and have not included them in the fits. 
%Since our J-band estimates of the radius have large 
%uncertainties, we have also used optical sizes---these are based on Palomar 
%Sky Survey images, and represent roughly the 25 mag arcsec$^{-2}$ B-band 
%isophotal size.

\begin{figure}
\plotone{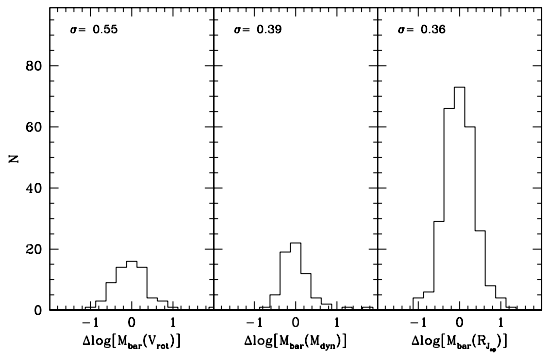}
\caption{The scatter of the baryonic masses around the least squares fits to 
each of the quantities shown in Figure 6. 
The rms dispersion of the fit is noted at the top of each panel.}
\label{fig:rathist}
\end{figure}

It is apparent by eye that the scatter is significantly smaller for galaxy radius 
than for other more traditional estimators of mass, although when we remove the
more face-on galaxies from our calculation, the difference between the dynamical
mass and J-band size is greatly reduced. We averaged forward and 
backward linear least squares fits to these data and display the distribution
around the fit and the calculated sigma in Figure \ref{fig:rathist}.
Thus, the size of a galaxy makes a good predictor of the mass present. While 
galaxy models do indicate that size might be the third parameter in a spiral 
galaxy fundamental plane relationship (Shen, Mo, \& Shu 2002), the physics 
behind this correlation is unclear.

%It is particularly interesting that the scatter is so small for the optical 
%size estimates, which, being based on rough estimates from optical photographs, 
%might be expected to have a high level of uncertainty. The small scatter may be
%an indication that the optical sizes are better determined than we had
%anticipated, or it may be a manifestation of size as a third parameter in the
%fundamental plane relationship.

\section{The Surface Density of Matter in Galaxies}

Despite the diverse range in galaxy properties discussed in the previous
sections, there is a tight correlation between HI mass and HI size (Rosenberg \&
Schneider 2003) as well as a good correlation between J-band luminosity and
J-band size. We examine the mass surface densities of gas, stars, and baryons
since the Schmidt Law associates the surface density of gas in a galaxy
with the star formation rate (Kennicutt 1998) and seems to imply a regulation 
mechanism between the gas and the star formation that might help explain the 
relationship between mass and size for these two populations.  

\begin{figure}
\plotone{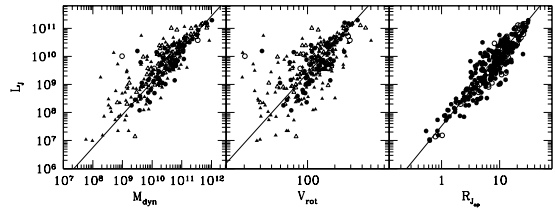}
\caption{The J-band luminosity for the RS galaxies (filled circles) and the SS
galaxies (open circles) plotted against the dynamical mass, rotational velocity,
and J-band size. The triangles in the dynamical mass and rotational velocity
plots indicate the galaxies with b/a$\ge$0.4. The solid 
lines show the forward and backward least squares fits to the data excluding
galaxies with b/a$\ge$0.4.}
\label{fig:lrcomp}
\end{figure}

Figure \ref{fig:lrcomp} shows the
relationship between luminosity and dynamical mass, rotational velocity, and
J-band size for these galaxies. Forward and backward linear least squares fits 
to these data are plotted. The standard deviation around these fits are 0.49,
0.69, and 0.37 respectively, providing an indication of the significance of the 
R$_J$ versus L$_J$ correlation. For the dynamical mass and rotational velocity
correlations we have plotted the galaxies with axis ratios greater then 0.4 as
circles (galaxies with axis ratios of 0.4 or less are plotted as triangles). We
have only included those galaxies with axis ratios greater than 0.4 in these two
fits. 

The relationship between J-band luminosity and J-band size is tighter than the 
Tully-Fisher relationship (Tully \& Fisher 1977) for this sample which includes 
irregulars, interacting systems, and dwarfs as well as spirals. Unfortunately, 
since luminosity and size scale the same way with 
distance, the tight correlation does not offer any possibilities for distance 
estimation,

The Tully-Fisher relation (Tully \& Fisher 1977) has provided an important test
of galaxy formation simulations and has been very hard to reproduce in detail;
most of the galaxies produced in simulations lose too much of their angular
momentum without large amounts of ad hoc energy injection into the system.
Governato et al. (2004) have managed to produce one galaxy with the correct
dynamical properties by going to higher resolution in their simulations, but
still has an excess of massive satellites and a paucity of cold gas. Even as
this work shows a success in producing the dynamical properties of a galaxy it
highlights how much is left to be done. The Tully-Fisher relation 
seems to indicate that there is a physical connection between the
dark matter component and the baryonic component in spiral galaxies.

For these
HI-selected galaxies, many of which are dwarfs or irregulars rather than spirals, 
we find less of a correlation between V$_{rot}$ or M$_{dyn}$ and M$_{bar}$ 
than is usually found for spiral galaxies (e.g., Giovanelli et al. 1997) culling
out the systems with axis ratios greater than or equal to 0.4. 
However, even for this disparate group of galaxies, the average surface density 
of gas (Rosenberg \& Schneider 2003) has a very small range relative to the
variation in the stellar surface density. The surface density of the gas is
determined as the HI mass divided by the HI area defined by the size at 2$\times
10^{20}$ cm$^{-2}$. 

\begin{figure}
\plotone{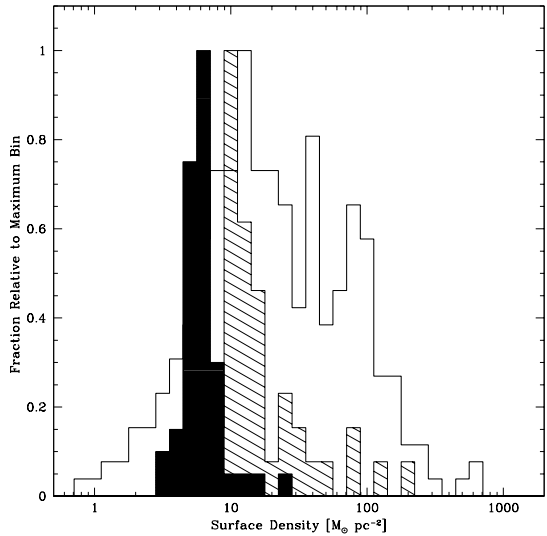}
\caption{The distribution of mass surface densities of gas,stars, and baryons. 
The gas surface densities, indicated by the dark-shaded histogram, come from the 
RS sample galaxies with measured HI sizes. The open histogram
shows the stellar mass surface densities of the RS and SS galaxies
using the J-band luminosity to estimate stellar mass, and the J-band sizes. The 
total baryonic mass 
surface density (cross-hatched histogram) is estimated using 
the sum of gas and stars and the optical size of each galaxy.} 
\label{fig:sd}
\end{figure}

Figure \ref{fig:sd} shows the HI surface densities for the the RS galaxies 
with measured HI sizes. Rosenberg \& Schneider (2003) showed that the 
average surface density of the gas in these galaxies was nearly constant, and
this result is evident in the small range in mean HI surface densities: the
values range from 3 to 24 M$_{\odot}$ pc$^{-2}$ with only 4 of the 50 galaxies
having values above 10. We do not discuss the surface densities of the SS sample 
because HI sizes are not available for these galaxies. Alternatively, the
stellar surface densities or the RS and SS sample cover a broad distribution
ranging from 0.8 to 660 M$_{\odot}$ pc$^{-2}$. The notable feature of 
this plot is that the dispersion in the HI surface density is
only 0.46 while the dispersion around the stellar mass surface density is 4.5.
The baryonic mass surface density is the combination of the HI and stellar
masses as given in Equation 2 and has a standard deviation around the mean
surface density (averaged over the HI size) of 5.8. These mass surface densities 
may provide an additional constraint for the simulations.

The larger dispersion in the stellar mass surface density of the RS and SS 
samples is, at least in part, due to the large uncertainties 
in the J-band sizes. Also, since the diameters were measured at a high surface 
brightness, the mean surface densities are higher than they would have been had 
we measured out to an equivalently low surface density as the gas. The median 
value for the distribution is 19.3 M$_{\odot}$ pc$^{-2}$ relative to 5.9 
M$_{\odot}$ pc$^{-2}$ for the HI mass density. We note that the median
value of the mass surface density is generally higher for the stars than for the 
gas. This corresponds well with the observation that many gas-rich galaxies have 
gas distributions extending well beyond the optical (or near infrared) light 
distributions. However, we do note that
there are still some systems in the sample that have very low stellar mass
surface densities providing an indication that we have detected some very low 
surface brightness systems.

\section {The Morphology of HI-Selected Galaxies}

Figures \ref{fig:adbs1} and \ref{fig:as1} in the Appendix show the 2MASS J-band 
images for the RS and SS galaxies with measured values of baryonic mass, 
respectively. The galaxies have been placed in order of their baryonic mass 
from the highest baryonic mass systems to the lowest (within each survey). 
Figures \ref{fig:adbs2} and \ref{fig:as2} show the galaxies for which we do not 
measure baryonic mass. 

The images of these galaxies illustrate the diverse population as discussed in 
\S 3. The sample covers the range from nearby bright spiral galaxies like RS
189, NGC 4565, to bulge-less low surface
brightness smudges like RS 184 (Figure \ref{fig:adbs1} in the middle of the
second to last line on the second page). There are also early type galaxies in
the sample like SS 19 (Figure \ref{fig:as1} in the middle of the last line)
which is NGC 7712, an elliptical galaxy (Scodeggio et al.
1995) and close interacting pairs like RS 50 (second to last image in Figure 
\ref{fig:adbs2}).

The SS98 optical images
of the SS sample indicate that a substantial fraction of the galaxies are low
surface brightness. As discussed in \S 3, these 2MASS data also indicate that
there is a substantial population of low surface brightness galaxies among the
RS and SS samples even though the surface brightness can not be directly
compared with the usual low surface brightness definition. Comparing the 2MASS 
images with the SS98 optical images, we
find that nearly all of the low surface brightness galaxies are not detected or
are barely detected in the 2MASS data. Inevitably this is because 2MASS is a
shallow survey, but there is no evidence for bulges that are 
bright in the infrared but not in the optical as was found for low surface 
brightness galaxies by Galaz et al. (2002).  

Visual inspection of the RS galaxy images seems to indicate that most of the 
highest baryonic mass galaxies are spirals with prominent bulge components 
while the lower
baryonic mass galaxies show more variation galaxy type and bulge size down to
galaxies like RS 184 where the brightness distribution appears fairly flat
across the J-band detected stellar disk. This same trend is not apparent for the
SS galaxies, but the sample size is significantly smaller which might account
for the difference.

\section{Summary}

We have examined the stellar properties of two HI-selected galaxy samples and
found that they show a large range of stellar properties. We find that the
galaxies cover a wide range in mass-to-luminosity ratio and the ratio is 
uncorrelated with the HI mass of the system. The range suggests that star 
formation does not proceed uniformly in all galaxies, some of the galaxies have 
funneled a large fraction of their gas mass into stars by the present day while 
there are other galaxies that are still largely dominated by their gas content. 
Because of these differences, one cannot infer the gas properties of a galaxy 
from its stellar properties and vice versa. The different proportions of gas and
stars in these galaxies may be providing us a glimpse of galaxies in different
stages of evolution.

Despite the wide range of gas-to-star ratios for galaxies in
these samples, there are some surprising correlations.
There is a small range in the average HI surface 
density in these galaxies. While the average stellar surface density distribution 
is not as tight, it is also a fairly narrow distribution given all of the
inaccuracies in its measurement. The baryonic matter appears
to average to about one quarter of the dynamical mass (within the optical dimensions of
the galaxy) across a wide range of galaxy types.

Most notably, the size of the stellar disk appears to be a very good predictor 
of the total baryonic content of galaxies. The physical causes of this 
correlation are not immediately apparent, but the correlation spans over three 
orders of magnitude in galaxy mass.

\acknowledgements

We would like to thank 2MASS for the effort that has gone into this catalog and 
for providing financial support for this research. We would particularly like
to thank Roc Cutri for all of the help in obtaining the full resolution images
that were needed for this paper to be possible. We would also like to thank 
the Arecibo and VLA staffs for their assistance with the HI observations. Thanks
also to Eric Bell for the information on J-band M/L ratios. We also appreciate
John Spitzak's work on the program to display and print the galaxy images.
Thanks also go to the anonymous referee for helpful suggestions. JLR
acknowledges support from the National Science Foundation under grant 
AST-0302049. The Digitized Sky Surveys were produced at the Space Telescope 
Science Institute under U.S. Government grant NAG W-2166. The images of these 
surveys are based on photographic data obtained using the Oschin Schmidt 
Telescope on Palomar Mountain and the UK Schmidt Telescope. The plates were 
processed into the present compressed digital form with the permission of these 
institutions.

\appendix

%\section{Appendix}

In this appendix we present the 2MASS data for the HI-selected samples. The data
include tables of the 2MASS size and magnitude values which are presented since
many of these values are measured from the images themselves or are derived from
the Version 2 extended source catalogs which are not generally available. The
information in Table 1, the ADBS data, is as follows: (1) the sequential listing 
in the RS catalog; (2) the name from the ADBS catalog; (3, 4) the RA and DEC 
from the HI catalog, (5,6) the RA and DEC from the 2MASS catalog if it was 
identified or of the 2MASS image
center, (7) the radius of the 21 mag arcsec$^{-2}$ isophote; (8) the b/a axis
ratio used to measure the magnitude within the 21 mag arcsec$^{-2}$ isophote;
(9) J 21 mag arcsec$^{-2}$ isophotal magnitude (J$_{21}$); (10) Error in
J$_{21}$, (11) the radius used for the aperture magnitude; (12) J-band aperture
magnitude; (13) error in the J-band aperture magnitude; (14) method used for
determining the magnitudes -- V2 and V3 refer to sources measured from the Version 
2 or 3 catalog, MI are sources measured off the images, and NM are sources for
which there was nothing on the 2MASS images to measure. The data in Table 2, the
SS data, are the same except that the SS catalog name and the sequential catalog
number are the same so all columns starting from the RA and DEC from the HI
catalog are shifted one to the left.

The images are presented as a quick look at the galaxies in the sample. The
objects have been placed in order of baryonic mass from the most massive to the
least massive. Objects in Figures \ref{fig:adbs2} and \ref{fig:as2} are sources 
for which a baryonic mass was not determined.

\begin{figure}
\plotone{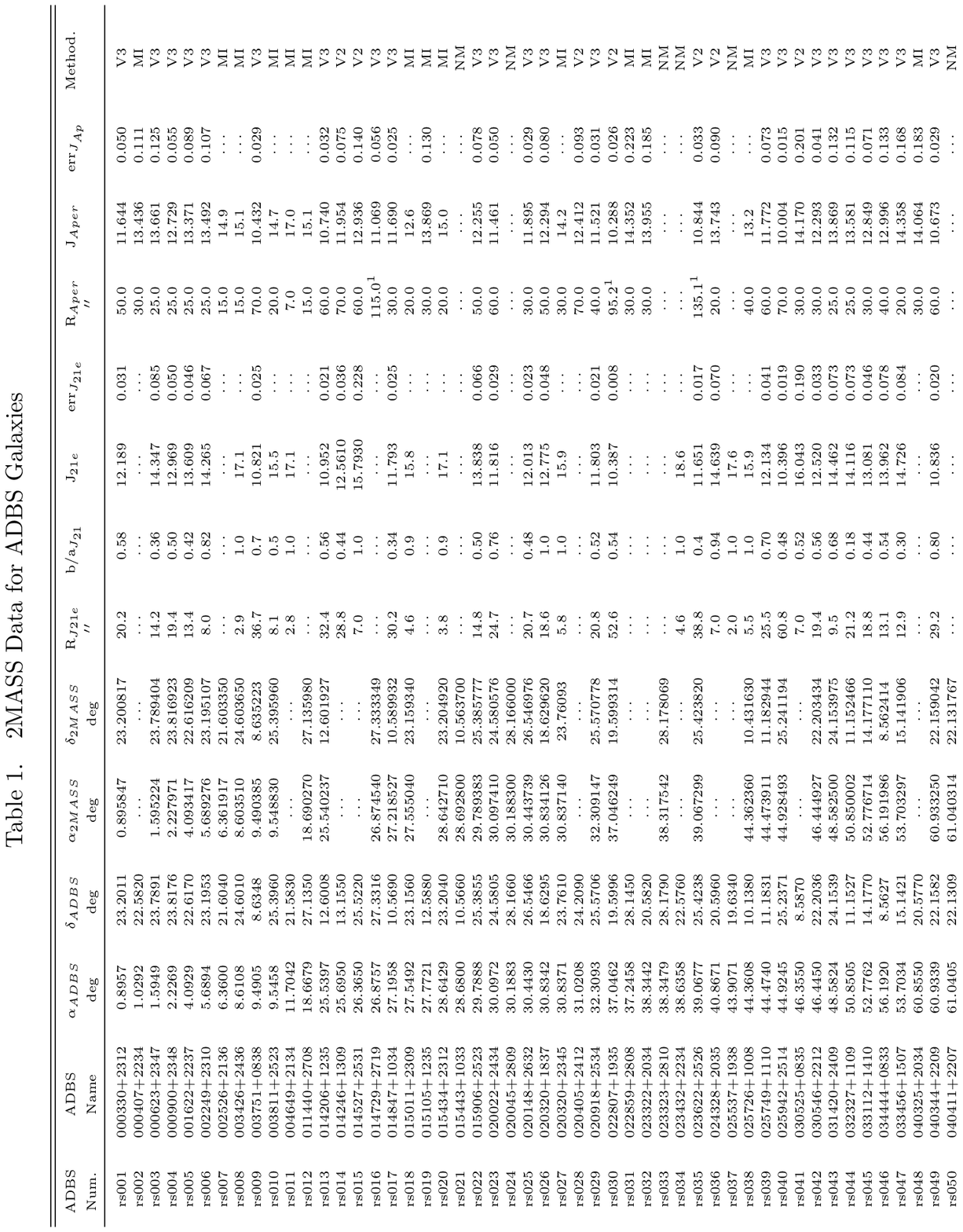}
\end{figure}

\begin{figure}
\plotone{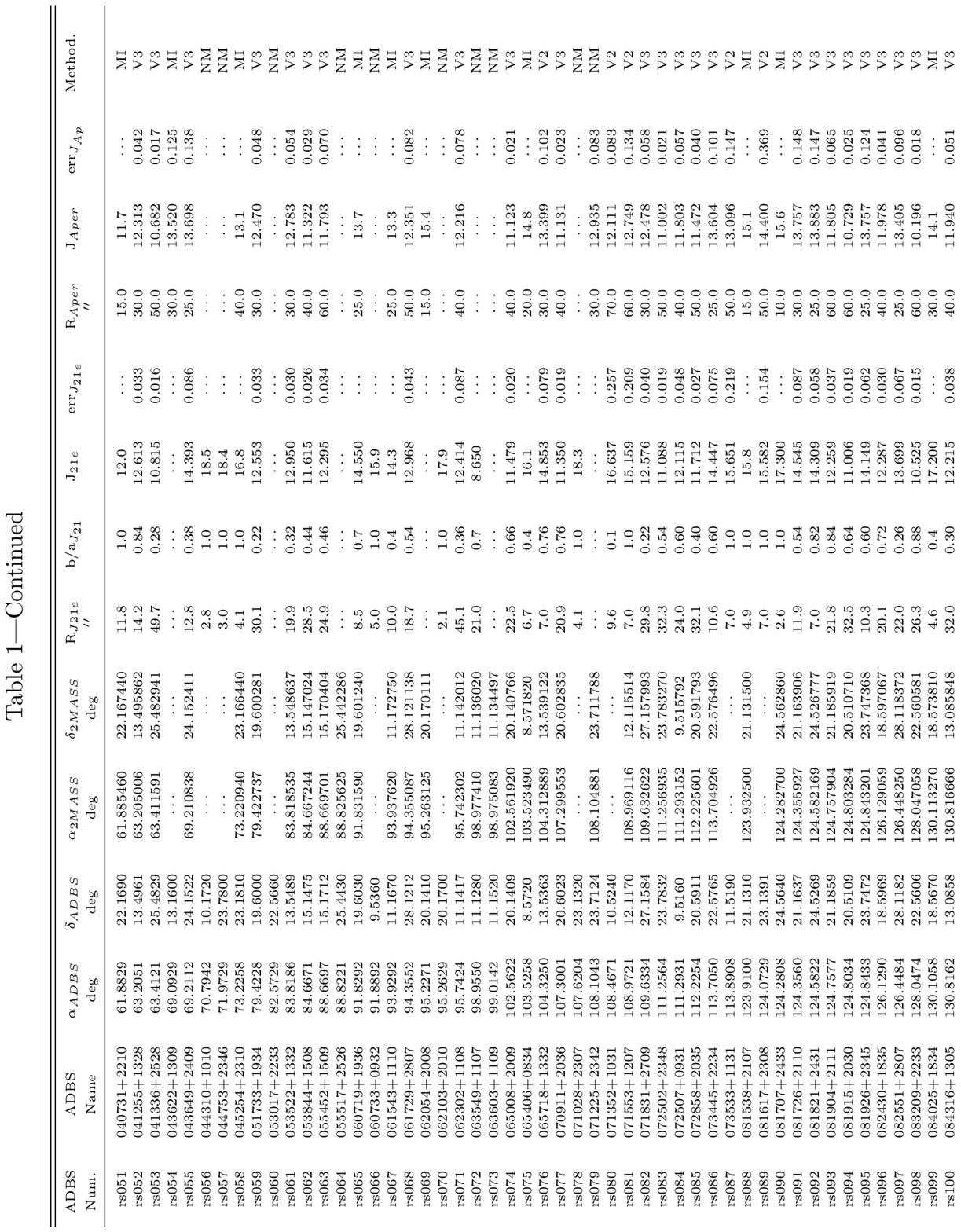}
\end{figure}

\begin{figure}
\plotone{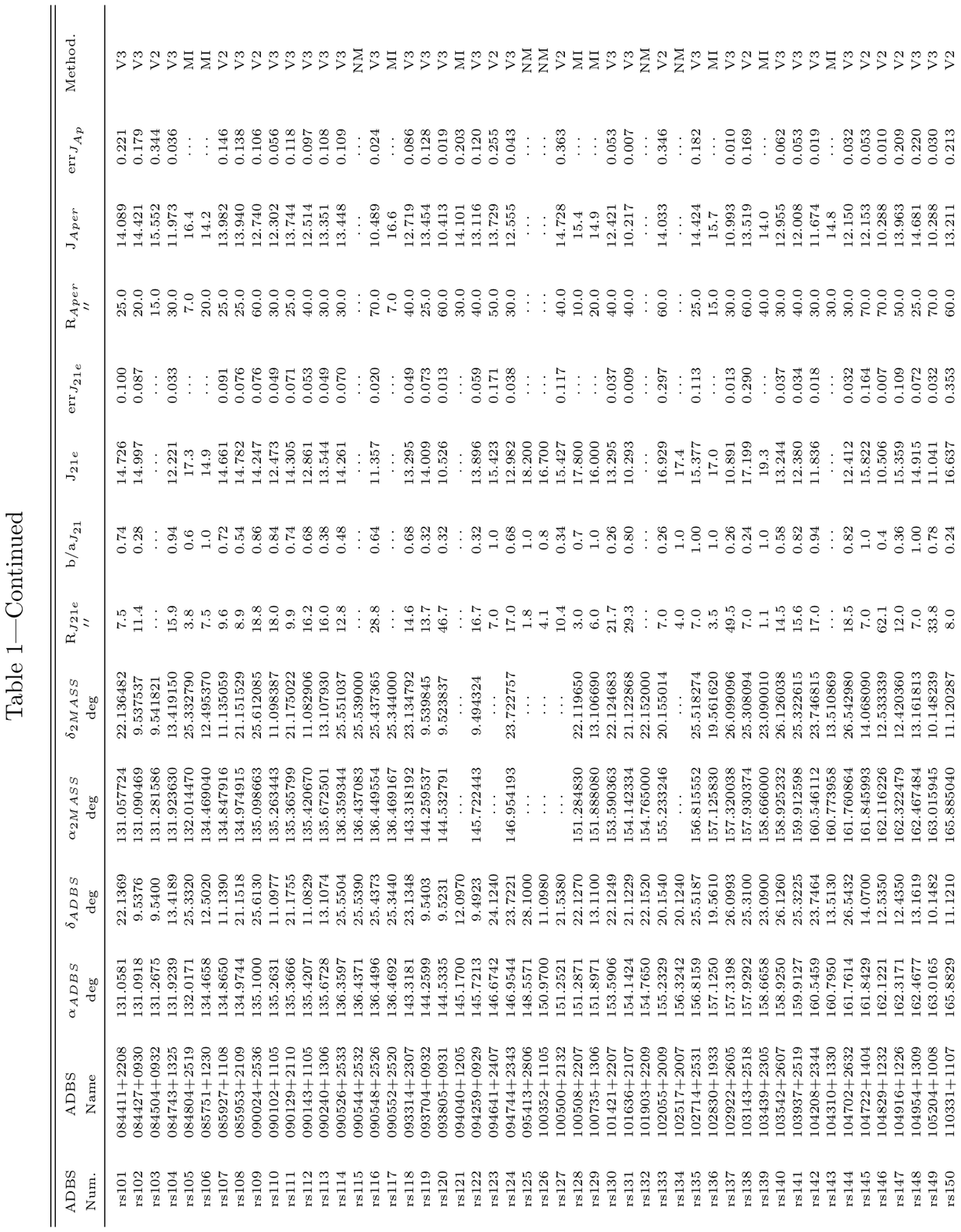}
\end{figure}

\begin{figure}
\plotone{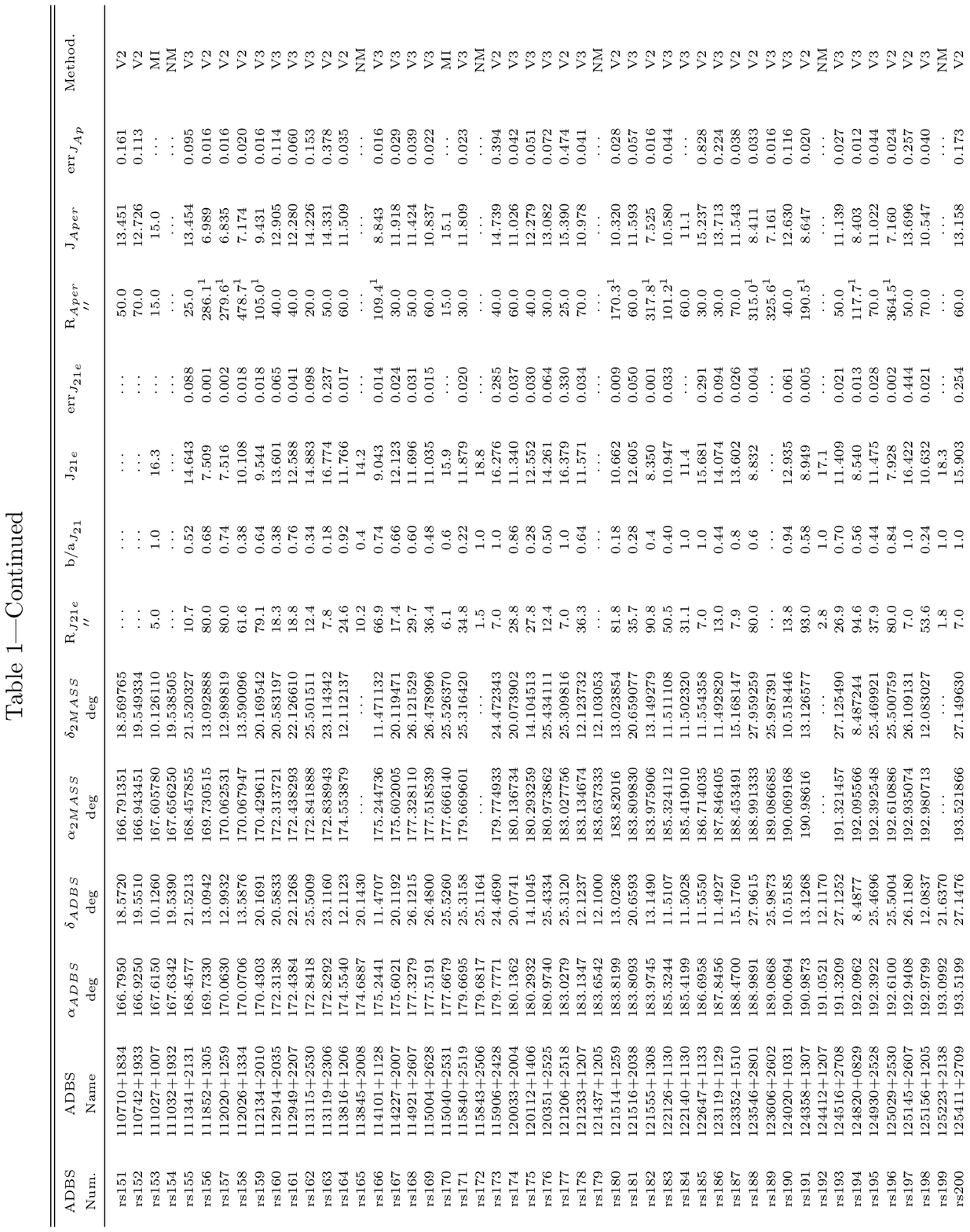}
\end{figure}

\begin{figure}
\plotone{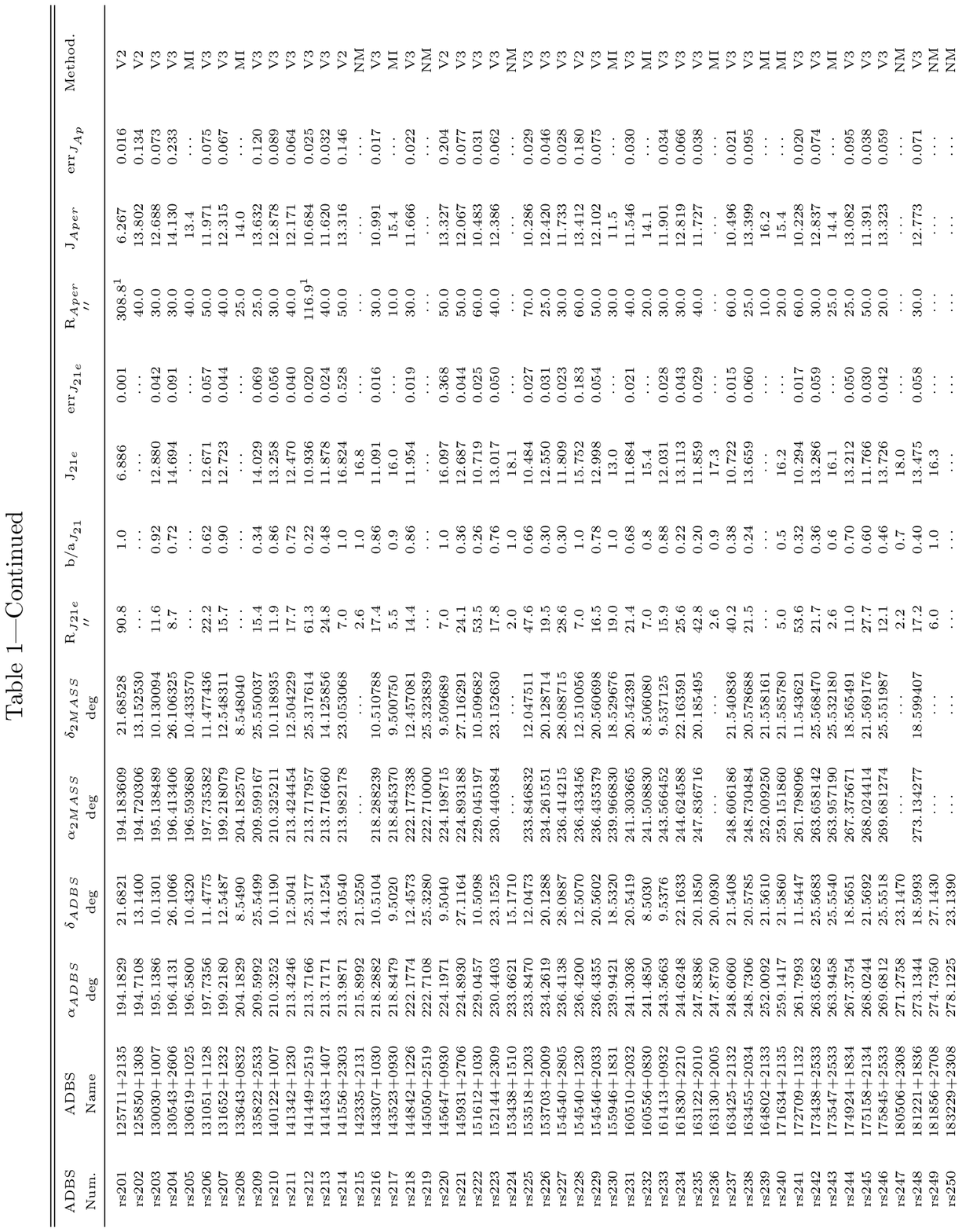}
\end{figure}

\begin{figure}
\plotone{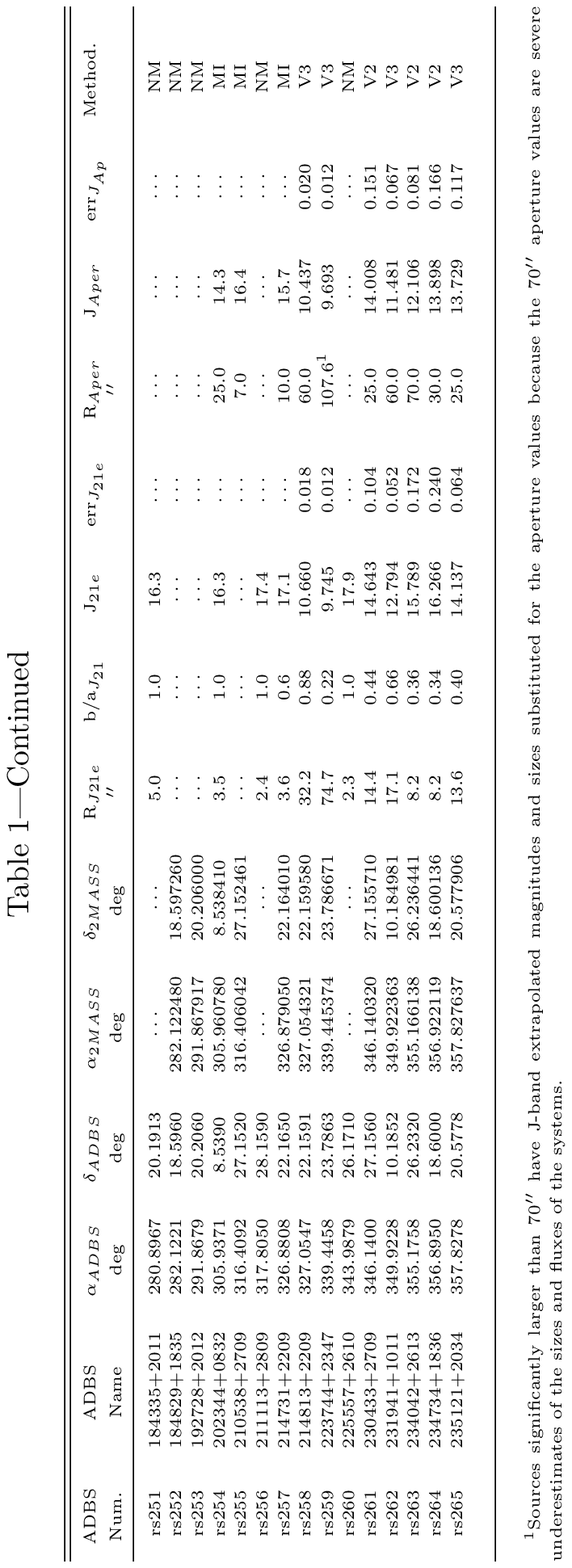}
\end{figure}

\begin{figure}
\plotone{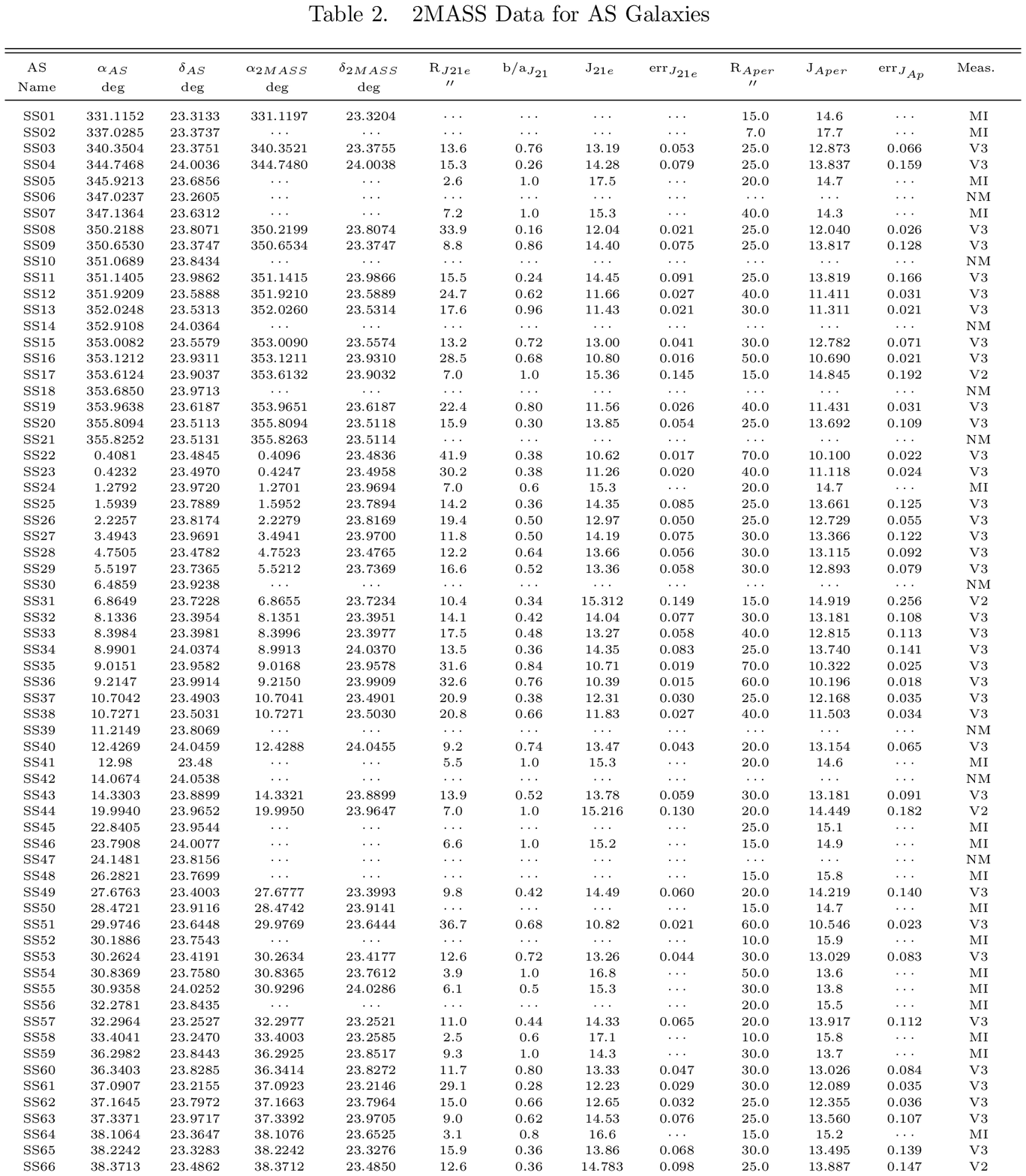}
\end{figure}

\begin{figure}
\plotone{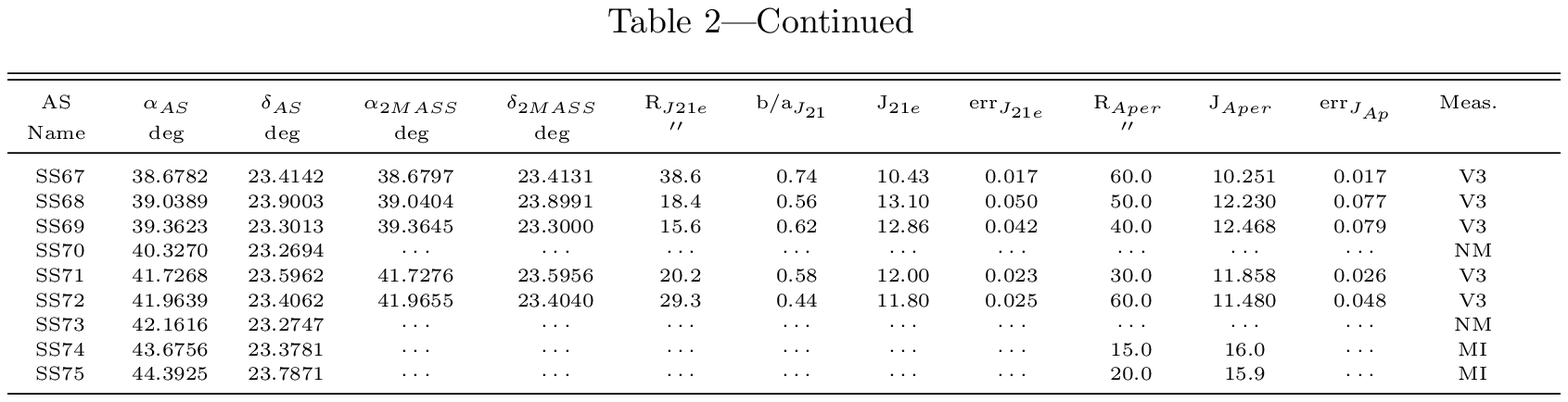}
\end{figure}

\begin{figure}
%\plotone{Rosenberg.figA1_comp.eps}
\caption{The ADBS galaxies in order of baryonic mass from the most massive to 
the least massive. (JPEG file included with download)}
\label{fig:adbs1}
\end{figure}

%\begin{figure}
%\plotone{Rosenberg.figA2_comp.eps}
%\end{figure}

%\begin{figure}
%\plotone{Rosenberg.figA3_comp.eps}
%\end{figure}

\begin{figure}
%\plotone{Rosenberg.figA4_comp.eps}
\caption{The ADBS galaxies for which baryonic mass was not able to be
determined. (JPEG file included with download)}
\label{fig:adbs2}
\end{figure}

\begin{figure}
%\plotone{Rosenberg.figA5_comp.eps}
\caption{The AS galaxies in order of baryonic mass from the most massive to 
the least massive. (JPEG file included with download)}
\label{fig:as1}
\end{figure}

\begin{figure}
%\plotone{Rosenberg.figA6_comp.eps}
\caption{The AS galaxies for which baryonic mass was not able to be
determined. (JPEG file included with download)}
\label{fig:as2}
\end{figure}

\end{document}